# The Determinants of Democracy Revisited: An Instrumental Variable Bayesian Model Averaging Approach

Sajad Rahimian

*Department of Economics, University of Saskatchewan, 9 Campus Drive, Saskatoon S7N 5A5, Canada*

Email address: sar291@usask.ca    ORCID iD: https://orcid.org/0000-0002-3187-2579

Biographical note

Sajad Rahimian is an M.A. candidate in economics at the University of Saskatchewan.

Acknowledgments

I thank Michael Jetter and Christopher F. Parmeter for sharing their codes. I am grateful to Jen Je Su for his helpful suggestions. Needless to say, all the remaining errors are mine.

# The Determinants of Democracy Revisited: An Instrumental Variable Bayesian Model Averaging Approach

Sajad Rahimian [*]

Abstract

Identifying the real causes of democracy is an ongoing debate. We contribute to the literature by examining the robustness of a comprehensive list of 42 potential determinants of democracy. We take a step forward and employ Instrumental Variable Bayesian Model Averaging (IVBMA) method to tackle endogeneity explicitly. Using the data of 111 countries, our IVBMA results mark arable land as the most persistent predictor of democracy with a posterior inclusion probability (PIP) of 0.961. Youth population (PIP: 0.893), life expectancy (PIP: 0.839), and GDP per capita (PIP: 0.758) are the next critical independent variables. In a subsample of 80 developing countries, in addition to arable land (PIP: 0.919), state fragility proves to be a significant determinant of democracy (PIP: 0.779).

Keywords: democracy; geography; endogeneity; model uncertainty

JEL classification: C11; D72; O11

---

[*] Department of Economics, University of Saskatchewan, 9 Campus Drive, Saskatoon S7N 5A5, Canada. Email address: sar291@usask.ca    ORCID iD: https://orcid.org/0000-0002-3187-2579

I. INTRODUCTION

Uruguay, Japan, and Germany are among the most democratic countries around the world, while Venezuela, Vietnam, and Belarus are host to authoritarian governments. Researchers have long struggled to find answers as to why democracy flourishes in one society and not in the other. It is a hard task, though. Manifested in the variegated theories in literature, economic (Lipset, 1959; Miletzki and Broten, 2017), institutional (Wyndow et al., 2013; Gerring et al., 2015), cultural (Inglehart and Welzel, 2005; Acemoglu et al., 2009; Grigoriadis, 2016), and geographical factors (Crenshaw, 1995; Engerman and Sokoloff, 2002; Elis et al., 2017) could matter in the democratization process.

The uncertainty about the driving forces of democracy has important policy implications. Presuming democracy as the best system of government currently available, what should be done to promote democracy at home or worldwide? Is economic liberalization, as Friedman (2009) suggested, an effective means? Alternatively, should the governments and international organizations put their emphasis primarily on economic development and obviating poverty? It might as well be the case that some relatively unchangeable cultural or geographical elements are the determining factors; thus, nations might be bound to inevitable destinies. In this paper, we aim to enhance our knowledge of the drivers of democracy or its absence. Notably, we strive to identify the main elements which promote or hinder democracy from a wide array of economic, institutional, cultural, and geographical factors.

Democracy literature is replete with heterogeneous results and has yet failed to pinpoint a set of robust democracy determinants. This could be attributed to two major problems, each hard to tackle. First, there are multiple democratization theories, emphasizing on various groups of variables as the underlying factors that matter the most to democracy. Some highlight economic factors, others underscore institutional, cultural, or geographical characteristics. The lack of an overarching theoretical framework has given way to model uncertainty in empirical democracy studies. It is not clear which control variables should generally be used in the econometric models. Second and probably the chief obstacle in the empirical literature is endogeneity, which largely stems from reverse causality. Democracy indeed affects some of its determinants. For instance, on the one hand, the conventional wisdom in literature confirms the positive effect of higher education on democracy (Glaeser et al., 2007). On the other hand, studies also provide support for the opposite cause-effect direction, that is, democratic governments extend education to more people (Gallego, 2010). Many studies have analyzed one or more democratization theories and by employing different variables and econometric methodologies (some take endogeneity into account, e.g. Jacobsen (2015), Gerring et al., (2018), and some neglect it, e.g. Gassebner et al., (2013), Oberdabernig et al., (2018)), they have produced a pool of results rich in contrasting and sometimes contradictory views.

To remove these two impediments, we take an unprecedented step forward. By utilizing Instrumental Variable Bayesian Model Averaging (IVBMA), introduced by Karl and Lenkoski (2012) and Koop et al. (2012), we simultaneously account for model uncertainty and endogeneity. In our analysis, we sort through a comprehensive set of 42 variables proposed by previous studies as candidate determinants of democracy. Our model space to be explored amounts to $2^{42}$ (4398046511104) different specifications. Moreover, our study takes advantage of the annual data from 1991-2010 and covers 111 countries, which constitute approximately 90% of the world population in 2010.

Among economic disciplines, growth researchers are pioneers in exploiting the capabilities of various Bayesian Model Averaging (BMA) techniques, including the IVBMA (Sala-i-Martin et al., 2004; Karl and Lenkoski, 2012). The IVBMA method has recently been used to address model uncertainty and endogeneity in corruption studies (Jetter and Parmeter, 2018). However, no researcher has applied IVBMA to topics in democracy literature. Finding proper instruments is a demanding challenge in the IVBMA context. Since the IVBMA technique is meant to encompass all relevant variables, it is tough to find instruments which satisfy the exclusion restriction (Jetter and Parmeter, 2018). We follow the lead in the closely related areas of economic research, e.g. growth and corruption studies, and instrument the possibly endogenous explanatory variables by their lagged values (Leon-Gonzalez and Vinayagathasan, 2015; Reed, 2015; Jetter and Parmeter, 2018).

Our dependent and explanatory variables are averaged values of annual observations between 2001 and 2010[1]. We instrument each possibly endogenous variable with its corresponding lagged value[2] (i.e. its averaged annual value from 1991 to 2000)[3].

There are stark differences between the results derived from our BMA and IVBMA models and each tag a unique group of variables as essential determinants of democracy. This indicates the influential impact of addressing endogeneity in the empirical analyses of democracy. While in the BMA model, cultural elements emerge as dominant factors, the IVBMA model puts its emphasis mostly on geographical and economic characteristics. More precisely, BMA identifies the Muslim population, socialist legal origin, and Gini index as the underlying causes of democracy, while IVBMA points to arable land, youth population, life expectancy, GDP per capita, and state fragility.

The rest of the article is organized as follows. Section 2 describes the data and reviews the literature. Section 3 discusses the research methodology. Section 4 presents the BMA and IVBMA results. Section 5 concludes.

II. POTENTIAL DETERMINANTS OF DEMOCRACY

In this section, we provide a general review of our data in the first subsection. We then proceed by introducing our index of democracy. Finally, we briefly survey the literature and introduce variables likely to be antecedent to democracy. Before starting, we first take a look at some of the recent empirical studies on the determinants of democracy in Table 1.

[Table 1]

---

1. The only exception is the *Military leader* variable which is computed based on annual data from 2001-2008.
2. We refer readers to Angrist et al. (1996), Imbens (2014), and Athey and Imbens (2017) for thorough discussions about causal inference and instrumental variables methodology.
3. There is one irregularity: the *State fragility* variable is instrumented using annual data from 1995-2000.

*2.1. Data*

Our data is structured as cross-section. We collected the data of 43 variables (one dependent variable and 42 independents) from 11 different sources at the country level. For all the variables, we averaged the annual data from 2001-2010. However, there is an exception: the *Military leader* variable is calculated based on annual observations from 2001-2008. We take our cue from Jetter and Ramírez Hassan (2015) and Jetter and Parmeter (2018) and average the data over ten years to alleviate concerns about possible measurement errors and business cycles. Our sample consists of 111 countries and covers approximately 90% of the world population in 2010. Variable descriptions and sources are given in Table 5. It should be mentioned that there is a trade-off between the number of variables and data availability. Data restrictions were significant burdens on our sample size and time period and we had to maintain an elusive balance between these aspects.

We should be cautious about a possible limitation of our study. Our sample might suffer from underrepresenting authoritarian states. A number of countries, mostly in MENA and Sub-Saharan Africa which have autocratic governments such as Saudi Arabia and the Democratic Republic of the Congo, are not included in our study. One should keep these limitations in mind when interpreting our results.

*2.2. Democracy index*

Democracy measurement has proved difficult in the social sciences. This is due to several reasons. First, definitions vary considerably. The competitiveness of electoral systems plays a central role in minimalist definitions (Alvarez et al., 1996; Przeworski et al., 2000). Moreover, social and economic features are incorporated into broader conceptualizations (Paxton, 2000; Freedom House, 2019). Second, democracy is closely related to other components of the institutional development (e.g. control of corruption, regulatory efficiency) (La porta et al., 1999; Papaioannou and Siourounis, 2008). Third, there is no consensus among scholars as to whether democracy is a categorical, continuous or a hybrid phenomenon (Wyndow et al., 2013). Among various democracy measures, the Freedom House (FH), Polity IV, and Vanhanen indices are the most famous. The FH has been criticized on methodological grounds since its civil liberties and political rights indices capture socioeconomic developments and do not exclusively evaluate political progress (Papaioannou and Siourounis, 2008). In addition, FH measures are perceived to be prone to severe measurement errors. The Polity IV, to some extent, suffers from similar setbacks (Munck and Verkuilen, 2002; Cheibub et al., 2010). In this article, we make use of Vanhanen's (2016) Index of Democratization. This index is built based on observable events and therefore is less likely to be biased by subjective judgments. Additionally, its construction draws on Dahl's (1971) prominent definition that recognizes electoral contestation and participation as the two sole central pillars of democracy. Moreover, Vanhanen's (2016) Index of Democratization has a competitive edge over other indices, because it includes participation; an important attribute of democracy that many other indices, e.g. Polity IV, have omitted (Munck and Verkuilen, 2002).

Figure 1 depicts the average democracy scores of all the 111 sample countries from 2001-2010. The highly democratic countries are illustrated in blue or green, whereas the authoritarian regimes are painted in red. See Table 6 for the full list of countries by region and their corresponding average democracy scores during 2001-2010.

[Figure 1]

In the next four subsections, we will introduce our variables and relevant studies. An exhaustive literature review is beyond the scope of this paper and because of space limitations, we try to keep our discussion as succinct as possible. Our variable classification, although based on literature, is subjective by its nature and some variables could belong to more than a group. Finally, it should be noted that our estimations are entirely independent of these categorizations. Table 5 gives information on the variables' descriptions and their respective categories.

## 2.3. Economic determinants

We start our literature review with economic determinants. According to the proponents of the modernization school of thought, economic development is an essential prerequisite for democracy (Lipset, 1959; Wucherpfennig and Deutsch, 2009). In the eyes of Lipset (1995) democratic systems need legitimacy in order to be stable, and legitimacy is acquired mainly through economic efficacy. Therefore, economic development is necessary for sustainable democracy (Miletzki and Broten, 2017). However, some scholars have challenged the validity of this theory (Acemoglu et al., 2009). In alignment with previous studies, we include GDP per capita, urbanization rate, secondary and primary education, infant mortality, agricultural employment (Lipset, 1959, 1995; Barro, 1999; Nieswiadomy and Strazicich, 2004), FDI (Escriba-Folch, 2017), life expectancy (Jacobsen, 2015), and Gini index (Savoia et al., 2010) variables in our models.

Economic freedom is another critical issue in the literature. Friedman (2009, p. 9) states: 'competitive capitalism also promotes political freedom because it separates economic power from political power and in this way enables the one to offset the other.' Accordingly, economic liberalization can spur democracy (Eichengreen and Leblang, 2008). There are studies which cast doubt on this theory (Li and Reuveny, 2003; Pond, 2018). To address these ideas we add economic globalization to our variables list. Moreover, as emphasized by Eichengreen and Leblang (2008, p. 289): 'the exchange of goods and services is a conduit for the exchange of ideas, and a more diverse stock of ideas encourages political competition.' Spilimbergo's article (2009) reveals that those international students who have studied in democratic countries nurture democracy in their home countries. On the contrary, Arif and Hall (2019) find no robust relationship between international travel and institutional quality in travelers' home countries. We enter the social globalization variable into our models to proxy for exchange in ideas and beliefs and to see which of the above discussions holds true.

Another theory puts natural resources wealth at the heart of its analysis. Natural resources abundance decreases the propensity for democracy (Ross, 2001; Andersen and Ross, 2014). Natural resources wealth could help governments suppress dissent and democratization by enabling them to use patronage, low tax rates, and obstructing the formation of non-government social groups (Ross, 2001). Contrary to these claims, Liou and Musgrave (2014) find that resources income might contribute to democratization in some situations. In our analysis, natural resources and fuel exports variables capture the effects of natural resources endowment on democracy.

*2.4. Institutional determinants*

As indicated by Table 5, institutional factors constitute another important group of our regressors. We start by presenting the population. The prevailing classical viewpoint affirms that democratic conduct and the size of polity are inversely related (Remmer, 2010). Smaller populations tend to have a stronger sense of community and this fosters civic engagement and decreases the cost of political activities. Moreover, individuals might perceive their votes to be more decisive and powerful in small communities and this will act as an added incentive for them to participate in electoral processes (Remmer, 2010). The opposite point of view, that the scale of the electorate and democracy could go hand in hand, has its supporters (Gerring et al., 2015). In addition to the sheer size of the population, Cincotta (2008a, 2008b) and Weber (2013) point out the implications that the age structure of the population could have for democracy. Their works unveil that the likelihood of a stable democracy diminishes as the share of young people in the population increases. These debates encourage us to add population and youth population variables to our analysis.

 Female empowerment is another important component of institutional category which could pave the way for democracy (Fish, 2002; Wyndow et al., 2013). Falling fertility rates could emancipate women from much of their childrearing and household activities, and provide women with more opportunities to further their education and participate in the workforce. These will perhaps increase women's awareness of the severity of discriminations against them and motivate them to push for gender equality and a type of political regime that is more responsive to their demands (Wyndow et al., 2013). We include fertility rate and female labor force variables to gauge the tenacity of these dynamics.

 Political stability is the last concept of interest in this category and is believed to be crucial for enduring democracy. Political stability emanates from a government's legitimacy and effectiveness and the extent to which elites are obedient to civil settings to take political power and their abstinence to resort to violence (Lipset, 1959, 1995; Epstein et al., 2006; Berg-

Schlosser, 2008). In this study, state fragility and military leader variables are our proxies for political stability.

*2.5. Cultural determinants*

Apart from economic and institutional predictors, a substantial segment of literature views cultural variables as the pivotal determinants of democracy. The legacy of Western colonizers has been the subject of numerous studies (Owolabi, 2015; Lee and Paine, 2019). Several articles have documented the effects of the Western powers' legal systems and ideology on their former colonies. Specifically, the British common law is regarded as more democracy and economic development friendly since it imposes stronger restrictions on the executives, better protects property rights and encourages free trade, compared to other legal systems (Olsson, 2009). Furthermore, Acemoglu et al. (2001, 2009) demonstrate that colonizers' strategies, whether Western settlers implemented despotic political systems and extractive institutions or established liberal traditions, have had persistent effects on the political and economic development paths of former colonies. To evaluate these arguments and represent countries' colonial heritage, we add nine dummy variables to our econometric models, under the "Colonial heritage" title in Tables 4 and 5.

Fractionalization and religion are other cultural elements which affect the likelihood of democracy. A widely held belief is that diverse social identities are inimical to democracy, but not everyone agrees with this assertion (Fish and Brooks, 2004; Gerring et al., 2018). Fragmentation might lower the social capital, cause conflict, or provoke rivalries among different sections of society to seize the power and hold on to it discretely (i Miquel, 2007; Gerring et al., 2018). Among the religions, Islam is of particular interest to scholars (Lewis, 1996; Huntington, 2000). Many studies have found evidence that proxies of Islam are negatively associated with democracy (Barro, 1999; Fish, 2002; Oberdabernig et al., 2018). Barro (1999) denotes the strong relationship between church and state in Islam as a plausible explanation. Fish (2002) opines that women subordination in Muslim countries is the main culprit for the democratic deficit. This should not be understood as if the status quo is eternal. Ross (2001, p. 339) reminds that: 'until the "third wave" of democratization began in the mid-1970s, democracy and Catholicism were negatively correlated.' The negative correlation between Islam and democracy might as well change in the future. We add five variables to our models, tagged as "Fractionalization and religion" in Tables 4 and 5, to test the robustness of these arguments.

Lastly, we wish to underscore the importance of citizens' cultural values in determining regime type as advocated by Inglehart and Welzel (2005). Self-expression values that encompass various aspirations such as civic participation, tolerance of others' liberties, human autonomy, generalized trust, and life satisfaction and happiness could influence the emergence and survival of democracy. However, like other democratization theories, Inglehart and Welzel's (2005) theory has not been immune to criticism (Dahlum and Knutsen, 2017a, 2017b). Unfortunately, we have not been able to incorporate these ideas into our empirical analysis due to severe data limitations. Data on emancipative values are drawn from World Values Survey waves and are available for 60 countries in the most recent wave (i.e. 2010-2014) which dramatically falls short from our sample of 111 countries.

2.6. Geographical determinants

An assessment of democracy determinants is not complete without appreciating the role of geographic features. Table 5 introduces our geographical explanatory variables. In their seminal work, Engerman and Sokoloff (2002) give a historical account of institutional and economic development in the Americas. Their study illuminates how geographic conditions, e.g. climate, the abundance of land and soil quality, provided a level ground for elites to amass a disproportionate share of political and economic power in Latin America through rent-seeking activities. North America's distinctive geographic attributes yielded a different pattern of institutional and economic development. Other authors have scrutinized the institutional and economic consequences of geographic characteristics (Crenshaw, 1995; Easterly, 2002; Bernauer and Vatter, 2017). In their study of causes of democracy, Csordás and Ludwig (2011) find that democracy transcends borders and permeates neighboring regions. Our models include nine geographical variables, specified under the title "Geographical" in Tables 4 and 5.

III. METHODOLOGY

It is necessary and insightful to introduce a number of studies which analyzed the causal links from democracy to our variables. We start this section by doing so and then continue by elaborating on the BMA and IVBMA frameworks.

*3.1. Endogenous variables*

The objective of this section is to briefly clarify the intuition behind the choice of our endogenous variables.

- Endogenous economic variables
    
    Glaeser et al. (2004, p. 271) identify two approaches to stimulate growth and establish democracy. They write that one approach:
    
    'emphasizes the need to start with democracy and other checks on government as the mechanisms for securing property rights. With such political institutions in place, investment in human and physical capital, and therefore economic growth, are expected to follow.'
    
    Besides, democratic governments offer better health services to their citizens compared to their autocratic counterparties (Patterson, 2017). We believe democracy influences our measures of economic development including GDP per capita, urbanization rate, infant mortality, life expectancy, and agricultural employment. So we consider these variables as endogenous. Acemoglu and Robinson (2000, p. 1168) provide a historical perspective on the democracy-inequality nexus and state that: 'democratization in turn opens the way for redistribution and mass education, and reduces inequality.' Their study prompts us to add Gini index, secondary education and primary education to the endogenous variables list. Eichengreen and Leblang (2008) investigate the intricate relationship between globalization and democracy and show that the two variables positively affect each other. Lastly, multinational corporations face political risks and fear dramatic policy shifts in host countries. In this respect, democratic governments

are more credible due to the presence of various checks and balances. The hospitable conditions in democratic systems attract FDI (Jensen, 2003). Based on these discussions, we perceive the economic globalization, social globalization, and FDI as endogenous.

- Endogenous institutional variables

    Democracy enhances female empowerment. Democracy lowers fertility rates and promotes job prospects for female labor force participants (Beer, 2009). Based on Dyson's (2013) narrative of demographic transitions, fertility rate and life expectancy are integral to changes in societies' age structures and populations. By influencing fertility rate and life expectancy, it is reasonable to assume that democracy has demographic effects. Thus, we consider the population, youth population, fertility rate, and female labor force variables as endogenous. Democratic governments guide citizens to solve conflicts of interests through voting and consensus, not violence. Therefore, democracy could contribute to stability (Blanco and Grier, 2009). With this in mind, we mark state fragility as our last endogenous variable.

*3.2. Tackling endogeneity*

We make use of Karl and Lenkoski's (2012) IVBMA procedure to deal with endogeneity and model uncertainty. Finding valid instruments is a major difficulty in the IVBMA context. Generally, geographical attributes or colonial heritages are the typical instruments in literature. Yet, these factors are amongst our independent variables, ruling out the possibility to use them as instrumental variables.

To deal with potential reverse causality, we take on a less adopted approach in the democracy literature and instrument each endogenous variable with its corresponding lagged value. More specifically, for each endogenous variable, we use its averaged value from 1991 to 2000 to instrument its respective averaged value from 2001 to 2010. We believe our approach is a reasonable one and works to tackle endogeneity because it gives consistent estimates of parameters, enables valid hypothesis testing (Reed, 2015), and produces informative outcomes (Bound et al., 1995; Conley et al., 2012).

Our decision to use lagged values of potentially endogenous variables as instruments is made upon four reasons:

(1) Researchers in other fields of empirical economic research have adopted this strategy. Examples are Temple (1999), Bhattacharyya and Hodler (2010), Schularick and Steger (2010), Arezki and Brückner (2011), Murtin (2013), Leon-Gonzalez and Vinayagathasan (2015), Mirestean and Tsangarides (2016), and Jetter and Parmeter (2018). Therefore, it is not an obscure practice in empirical economic research.

(2) We find instances of partially similar exercises in democracy literature. In his notable study of determinants of democracy, Barro (1999, p. 163) reports that:

> 'The empirical results turn out to be virtually the same if contemporaneous values of the Z variables are entered into eq.(1)[4] (i.e., if the lag T is set to zero), but lagged values of the Z variables and the lags of democracy are used as instruments.'

In another well-received study, Acemoglu et al. (2008) employ instrumental variables regressions to isolate the causal effect of income on democracy. To address endogeneity, they use past savings rate as an instrument for income per capita.

(3) A valid instrumental variable (IV) ought to have two essential qualities: first, it shows a high correlation with the corresponding endogenous explanatory variable; second, it fulfils the exclusion restriction[5]. Regarding the former characteristic, almost all of our instruments exhibit strong associations with their respective endogenous explanatory

---

4 . Barro (1999, p. 163) estimates the following equation:

$$\text{DEMOC}_{it} = \alpha_{0t} + \alpha_1 \text{DEMOC}_{i,t-T} + \alpha_2 \text{DEMOC}_{i,t-2T} + \alpha_3 Z_{i,t-T} + u_{it}$$

'where i is the country; t is the time period; T is a time lag, usually taken to be five years; DEMOC is the indicator for democracy - electoral rights or civil liberties; Z is a vector of variables, such as per capita gross domestic product and education, that influence the extent of democracy; and u is an error term.'

5 . Exclusion restriction stipulates that a credible instrument alters the outcome variable just through its effect on the endogenous independent variable.

variables. The correlation statistics, in descending order, are 0.99 for population, 0.99 for GDP per capita, 0.99 for urbanization rate, 0.99 for agricultural employment, 0.98 for social globalization, 0.98 for fertility rate, 0.97 for state fragility, 0.97 for infant mortality, 0.97 for female labor force, 0.97 for life expectancy, 0.94 for primary education, 0.93 for secondary education, 0.91 for Gini index, 0.81 for economic globalization, 0.8 for youth population, and 0.65 for FDI.

On the contrary, the second attribute (i.e. excludability) is quite difficult to satisfy and evaluate[6]. It is not hard to think of plausible scenarios where our instruments might exert an influence on the democracy index through channels other than the endogenous explanatory variables. But, obtaining insightful results might not be conditional on strict excludability of instrumental variables.

Bound et al. (1995) demonstrate that when instruments and endogenous variables are weakly correlated, even a slight deviation from the exclusion restriction could result in heavily biased estimates, but this is not the case when instruments and endogenous variables are strongly correlated. Conley et al. (2012, p. 261) emphasize this point and state that 'large deviations [from the exclusion restriction] may have only small influences on [estimate] precision when the instruments are strong [i.e. when they are highly correlated with endogenous explanatory variables].' Conley et al. (2012, p. 260) illustrate that when instruments are strongly correlated with endogenous variables, 'inference is informative even with a substantial relaxation of the exclusion restriction'. Since our instruments exhibit strong correlations with their respective endogenous

---

6 . We draw the readers' attention to Jiang (2017, p. 9):

'Perhaps, except for a godsend, it is challenging to find a valid instrument that could explain a significant portion of the variation in an endogenous independent variable of key interest. That's because such a powerful explanatory variable would, most likely, already be in the system, which means that it would, most likely, directly affect an outcome in most cases, thus violating the exclusion restriction.'

variables, we could be relatively confident that our results are informative even if the exclusion restriction is not completely fulfilled.

(4) Reed (2015) runs various simulations using different data generating processes and analyses the performance of lagged explanatory variables as instruments in 2SLS estimation procedures. His simulations unveil that using lagged independent variables as instruments would result in consistent estimates of parameters and enable valid hypothesis testing if the following two preconditions are met:

1. Lagged explanatory variables show high correlations with current explanatory variables.

2. Lagged explanatory variables do not belong in the primary equation of interest. In our application, this requirement translates that we have to be able to reasonably assume lagged explanatory variables (i.e. 1991-2000 averages) do not exert a direct influence on the subsequent democracy levels (i.e. 2001-2010 averages).

We previously showed that the first precondition is met properly. Considering the second requirement, since we have used decade-long averages in our empirical analysis, we could expect that democracy levels are mostly affected by variables during the same decade, not the previous one.

These deliberations and observations give us confidence that our instruments are, though not entirely but largely, reliable and have roots in literature. Nonetheless, we do not intend to dismiss the shortcomings of our instruments, particularly concerns over their compliance with the exclusion restriction. Instead, we believe that it is crucial to replicate our analysis with other sets of instrumental variables and scrutinize the robustness of our results in future research.

More broadly, it is necessary to touch on the limitations of the instrumental variables approach. Controlled randomized experiments remain the golden tools for unveiling causal relationships. Yet, randomized experiments are a luxury many researchers cannot afford. In most applications, researchers only have access to observational data. IV methodology is an effective econometric tool to make causal inferences from observational data, but it comes with its disadvantages. Most notably, strong assumptions of the IV approach are its main drawbacks. Finding a valid instrument that correlates well with the endogenous variable and satisfies the exclusion restriction is far from easy. Further, different instruments might yield different results as they could have heterogeneous effects on the endogenous variable (Sovey and Green, 2011; Becker, 2016). Moreover, IV methodology could not tell much about the network of relationships that might exist between the right-hand side variables of a regression model. Specifically, concerning this limitation, Data Acyclic Graphs (DAGs) could be very revealing and provide a possible avenue for future research in democracy literature.

Now, having said the merits and limitations of our instrumental variable approach, we continue by briefly outlining the principles of BMA and IVBMA frameworks.

*3.3. Bayesian Model Averaging (BMA)*

Researchers usually encounter two types of uncertainty in empirical studies: model and parameter uncertainty. Model uncertainty arises when literature offers numerous explanatory variables and researchers are not certain about the relevant set of variables to include in the right-hand side of their econometric models. Suppose that previous studies and theories have proposed *k* candidate explanatory variables. Then, there are $2^k$ possible combinations of these variables, i.e. $2^k$ models. Model uncertainty is a serious problem especially when literature suggests many potential determinants. A non-transparent and arbitrary variable selection procedure might result in omitted variable bias and yield erroneous conclusions and estimates that do not correctly describe the underlying data generating process. Parameter uncertainty refers to the ambiguity of parameters' values conditional on a given model. Researchers use a vast array of econometric methods to estimate the required parameters (Wintle et al., 2003; Arin and Braunfels, 2018).

Bayesian Model Averaging[7], introduced by the pioneering work of Raftery et al. (1997), has been widely used in economic analyses to tackle model uncertainty[8]. BMA also provides estimates of desired parameters. Thus, it helps to overcome parameter uncertainty as well.

---

7 . A competent introduction to BMA is available by Hoeting et al. (1999).

8 . Economic growth is a research area filled with Bayesian Model Averaging papers.

Suppose a linear regression model:

$$y = \beta_0 + \beta X + \varepsilon, \tag{1}$$

where y denotes the dependent variable and X represents K potential determinants. BMA estimates all possible combinations of explanatory variables, which are $2^K$ combinations and thus $2^K$ models. The posterior model probability (PMP) of a model given the data, $P(M_\gamma|D)$, gives the probability that model $M_\gamma$ is the true model. It is the ratio of the marginal likelihood of model $M_\gamma$ to the sum of marginal likelihoods of all possible models:

$$PMP_\gamma = P(M_\gamma|D) = P(D|M_\gamma)\, P(M_\gamma) / (\sum_{i=1}^{2^K} P(D|M_i)\, P(M_i)), \tag{2}$$

where

$$P(D|M_\gamma) = \int P(D|\beta_\gamma, M_\gamma)\, P(\beta_\gamma|M_\gamma)\, d\beta_\gamma, \tag{3}$$

and '$\beta_\gamma$ is the vector of parameters from model $M_\gamma$, $P(\beta_\gamma|M_\gamma)$ is a prior probability distribution assigned to the parameters of model $M_\gamma$, and $P(M_\gamma)$ is the prior probability that $M_\gamma$ is the true model' (Grechyna, 2016, p. 12). The posterior inclusion probability (PIP) of a particular variable x is the index of its importance (Zeugner, 2011), and is the sum of PMPs for all models which contained it:

$$P(\beta_x \neq 0|D) = \sum_{\gamma: \beta_x \in M_\gamma, \beta_x \neq 0} P(M_\gamma|D). \tag{4}$$

Also,

$$E(\beta_x|D) = \sum_{i=1}^{2^K} \hat{B}_i\, P(M_i|D), \tag{5}$$

Equation (5) indicates that the posterior mean of any particular variable x is the weighted sum of its coefficients ($\hat{B}_i$) across all models; where weights are the PMPs of models.

Equation (6) outlines the formula for the posterior variance of variable x:

$$\text{Var}(\beta_x|D) = \sum_{i=1}^{2^K}(\text{Var}(\beta_x|D, M_i) + \hat{B}_i^{\,2})P(M_i|D) - E(\beta_x|D)^2 \tag{6}$$

There are a number of advantages to the BMA method:

- 'If you torture the data enough, it will confess.' (Coase)
  One can obtain their desired outcomes by examining different model specifications and selecting the one which best fits their expectations. One of the advantages of the BMA method is that it refrains from forcing a predefined model to data and instead allows the data to decide the best model for describing the dependent variable. BMA does not restrict the data generating process to a specific model.
- BMA allows the researcher to enter their prior beliefs or information about the true model to the analysis. BMA combines these prior assumptions with the information derived from data and produces posterior results.
- Functional form uncertainty is another major concern in empirical studies. This problem usually relates to the functional specification of the regression model, whether the econometric model should be linear or non-linear. BMA techniques have improved significantly in recent years and there are BMA methods to model the non-linear relationships between variables of interest[9].
- BMA can be used in the panel data context. Furthermore, it can assist to overcome endogeneity problems[10].
- BMA is computationally efficient. It can handle very large model spaces by exploiting the capabilities of various Markov chain Monte Carlo methods.
- Results are straightforward and easy to interpret. Posterior probabilities could be compared across models or variables without difficulty.

*3.4. Instrumental Variable Bayesian Model Averaging (IVBMA)*

Karl and Lenkoski (2012) and Koop et al. (2012) developed the IVBMA method to explicitly address endogeneity in the BMA context. In this article, we closely follow the procedures

---

9 . See Shi (2016) and Balcombe and Fraser (2017) for examples.

10 . Two examples are Leon-Gonzalez and Montolio (2015) and Leon-Gonzalez and Vinayagathasan (2015).

proposed by Karl and Lenkoski (2012). Consider the classic, two-stage endogenous variable model:

$$Y = X\beta + W\gamma + \epsilon \tag{7}$$

$$X = Z\delta + W\tau + \eta \tag{8}$$

with

$$\begin{pmatrix} \epsilon_i \\ \eta_i \end{pmatrix} \sim N_2(0, \Sigma) \tag{9}$$

and

$$\Sigma = \begin{pmatrix} \sigma_{11} & \sigma_{12} \\ \sigma_{21} & \sigma_{22} \end{pmatrix}; \sigma_{12} = \sigma_{21} \neq 0 \tag{10}$$

while $Y$ denotes the vector of response variable, $X$ is the matrix of endogenous explanatory variables, $W$ denotes the matrix of exogenous explanatory variables, and $Z$ is the matrix of instrumental variables. $\epsilon$ and $\eta$ are homoscedastic and correlated error terms. Since cov $(\epsilon, \eta) = \sigma_{12} = \sigma_{21} \neq 0$, we have to estimate the equations (7) and (8) jointly to draw valid inferences about the parameters in the primary equation (7).

Karl and Lenkoski (2012) incorporate uncertainty into both equations (7) and (8). This implies that not only the best determinants of $Y$ are unclear (i.e. uncertainty in equation (7)), also the robust predictors of $X$ are ambiguous (i.e. uncertainty in equation (8)). Karl and Lenkoski (2012) introduce IVBMA to surmount these two types of uncertainties.

Their IVBMA procedure has two main phases. Suppose that a subset of $X$ and $W$ variables are chosen to describe $Y$ in equation (7). Also, assume a subgroup of $Z$ and $W$ covariates are entered in equation (8) to predict $X$. Given this starting current state, IVBMA works as follows:

1. A new model for the equation (7) is suggested that differs from the current model by only one variable. Karl and Lenkoski (2012) use a uniform prior on space of models meaning that possible models have equal chances to get selected. A modified type of Gibbs sampler called "MC3-within-Gibbs" is employed to explore the model space and suggest new models.

IVBMA uses a conditional Bayes factor (CBF) to compare the posterior probabilities of the current and suggested models and detect the one that has higher odds. If the suggested model is accepted then the parameters and error term of the equation (7) are updated.

2. The second step is similar to the first phase in spirit, except that it focuses on equation (8). In the second step, a new model for equation (8) is proposed that diverges from the current model by only one variable. A uniform prior on space of models is assumed, meaning equal selection chances for all possible models. Similar to phase one, IVBMA uses MC3-within-Gibbs sampler to explore the model space and move between models. Karl and Lenkoski (2012) use a conditional Bayes factor to compare the posterior probabilities of the current and proposed models and figure out the model with superior probability. If the proposed model is accepted then the parameters and error term of the equation (8) are updated. The estimated $X$ values enter the equation (7) and the IVBMA loop runs again.

By repeating this process numerous times, IVBMA moves toward models that perform better in predicting $Y$ and $X$. Meanwhile, IVBMA saves the sequence of encountered models for each of the equations (7) and (8). Since the second stage equation (i.e. equation (7)) is the primary equation of interest, the pertinent sequence of models could be used, based on equations (4) and (5), to calculate the PIPs and posterior means of explanatory variables.

IV. EMPIRICAL RESULTS

We use the *BMS* package developed by Zeugner and Feldkircher (2015) to execute the BMA model. The *ivbma* package from Lenkoski et al. (2014) carries out the IVBMA models. Both packages are available at the Comprehensive R Archive Network[11]. Throughout this section, it should be noted that our BMA and IVBMA methods assume a linear functional form when analysing the relationship between explanatory variables and the democracy index. Literature provides evidence of nonlinearities in the democratization process (Spaiser et al. 2014). Covering non-linear complexities in future research could be a valuable contribution to democracy literature.

---

11 . The ivbma package can be downloaded from https://cran.r-project.org/src/contrib/Archive/ivbma/

We run 3000000 iterations and discard the first 200000 draws as the burn-ins for each of our three models. Data and codes are available upon request. Throughout this section, we follow Kass and Raftery (1995) and Eicher et al. (2012) classification, who categorize PIPs as follows:  PIP > 0.99 presents decisive evidence, 0.95 < PIP < 0.99 indicates strong evidence, 0.75 < PIP < 0.95 provides positive evidence, and PIP < 0.75 is perceived as weak evidence. We have reported the BMA results in Table 2 to make comparisons easier between BMA and IVBMA outcomes. But we present the detailed description of BMA results in the Supplementary Information file since our main focus is on addressing the endogeneity and model uncertainty simultaneously through IVBMA models.

[Table 2]

*4.1. IVBMA results - full sample*

Columns 5-7 of Table 2 summarize the results from the second stage of our main IVBMA model, after instrumenting potentially endogenous explanatory variables with their lagged values. The first vivid point is that in striking contrast to the BMA estimation, our basic IVBMA model picks an utterly different set of variables as the robust regressors of democracy. To us, this indicates the dire repercussions of ignoring endogeneity and how overlooking this issue could mislead researchers into faulty conclusions. It is interesting to see how poorly Muslim population, socialist legal origin, and Gini index[12] variables are doing in the IVBMA context. The PIP of Gini index drops dramatically to 0.117, the PIP of socialist legal origin shrinks to 0.594, and the PIP of Muslim population decreases to 0.731, narrowly rendering it as a weak predictor of democracy. On the contrary, after we take care of endogeneity, some variables witness sharp rises in their PIPs. According to the main IVBMA results, arable land is the most persistent predictor of democracy with a PIP of 0.961 (note that arable land has a PIP as low as 0.423 in the BMA model). The youth population is the second robust determinant of democracy with a PIP of 0.893 (notice its low PIP of 0.244 in the BMA model). Life expectancy and GDP per capita are the subsequent important causes of democracy with PIPs of 0.839 and 0.758 respectively. Arable land, life expectancy, and GDP per capita possess positive posterior means. This means that their increase enhances democracy. The negative posterior mean of the youth population conveys that its increase weakens democracy.

---

12 . Based on the BMA model, these three variables have the highest PIPs among the potential democracy determinants, making them the fundamental predictors of democracy.

What is remarkable is the identification of arable land as the most influential predictor of democracy in the IVBMA model. To the extent of our knowledge, there are three studies by Hegre et al. (2012), Gassebner et al. (2013), and Oberdabernig et al. (2018) that are similar to our article in their objectives but differ considerably in methodology. All three emphasize the importance of addressing endogeneity and all fail to tackle it. Hegre et al. (2012) do not use arable land and the other two articles find it as an insignificant explanatory variable. Since our paper is the first article that systematically tackles both model uncertainty and endogeneity, it would not be surprising that it offers a somewhat unorthodox perspective.

The fairly unexpected emergence of arable land as the most robust determinant of democracy in the IVBMA estimation motivates us to expand on the role played by geographic conditions in the creation and solidification of democracy. In our study, arable land serves as a proxy for the Crenshaw's (1995) proto-modernization hypothesis[13]. The high PIP of arable land endorses Crenshaw's (1995) vision[14]. Many of Crenshaw's (1995) arguments revolve around *technoecology*. A concept which he defines as: 'a population's adaptation to social and biophysical environments through changes in sociospatial organization' Crenshaw (1995, p. 705). In his view, climate, the abundance of land and its ownership system, disease regime, geography, the social environment and previously developed technologies are elements of technoecology. Technoecology affects the size and density of population, division of labor, innovation and differentiation, agriculture production, economic surplus, nonfarm employment, commercialization, the advent of possessive individualism, and even colonial histories. Crenshaw (1995, p. 706) states that:

> 'the fact that Europeans actually encountered states in high-density societies illustrates how technoecological heritage allowed some societies to escape colonialism altogether (e.g., Japan, Thailand), and others to escape colonial status much more quickly than would otherwise have been the case.'

---

13 . Proto-modernity lays stress on a country's ecological heritage and its pre-industrial past. It contends that whether or not a country has a history of rich and advanced agrarian social organization, plays a critical role in defining its capacity to sustain a democratic form of government in the modern era.

14 . One might cast doubt as to whether arable land is a proper proxy for pre-industrial history or ecological legacies. While such criticism could have merits, we believe arable land is a suitable proxy. We are inspired by Crenshaw (1995), Gassebner et al. (2013), Oberdabernig et al. (2018), and other studies in democracy literature to select this variable. We leave this issue open for future investigations.

Briefly, favorable technoecological conditions furnished the basis for complex and developed agrarian social structures. The most democratic countries in modern ages are usually heirs to these rich and advanced agrarian societies.

Engerman and Sokoloff (2002) adroitly uncover the links between geographic conditions and political and economic development in the Americas. Take Brazil and the United States for example. In Brazil, climate and soil quality, among other factors, encouraged colonizers to grow hugely profitable crops such as sugar on large slave plantations. Massive numbers of slaves were imported to supply the required labor. These dynamics led to the creation of a strictly hierarchical society with an affluent and politically powerful minority and an impoverished and powerless majority. The United States has a different experience. Its lands were not suitable for the lucrative sugar business, but they were fertile for growing grains and livestock farming. Farms were smaller and usually owned by families. The demand for slave labor force was relatively limited and slaves constituted a smaller fraction of the population. A combination of climate, land ownership, agriculture and demography created a much more egalitarian society in the United States (Engerman and Sokoloff, 2002).

Acemoglu et al. (2001, 2002, 2009) demonstrate that events in critical historical junctures have enduring effects on countries' institutional and economic development paths. Geographic characteristics become decisive during these crucial periods. For instance, the climate and land features prompted European settlers to establish different institutions in former colonies. In places where the geographic conditions were not hospitable and colonizers suffered high mortality rates, they inclined to extractive and despotic social structures. In palatable environments, they tend more to build liberal institutions. Institutions tend to persist and economic and political institutions formed in distant past could partly survive to present time (Acemoglu et al., 2009).

Our findings regarding youth population, life expectancy and GDP per capita are quite standard and well documented in the literature. Having large proportions of young people is not good news for democracy. Weber (2013) discusses the implications of a young age structure for democracy in details. Based on his line of reasoning, diverse identities (e.g. religious and cultural) and social antagonisms exist in all societies, including democratic ones. A stable democracy solves interest and identity conflicts through negotiation and argumentation. Those who consider their opponents as enemies, reject compromise and are willing to use force to protect their interests or advance their cause are a threat to democracy. Due to identity confusion, the youth are more prone to stereotyping their beliefs and enemies and to over-identify with their favorite social groups. Intolerance, intergroup hate, and support for demagogues could be more rampant among the youth. Therefore, a relatively large youth cohort size could be dangerous for democracy. Weber's (2013) remarks are compatible with "youth bulge" theory, advocated by Cincotta (2008a, 2008b).

Life expectancy and GDP per capita are two of the flagship variables of modernization theory and have been extensively used in empirical studies (Epstein et al., 2006; Wucherpfennig and Deutsch, 2009; Jacobsen, 2015). Higher GDP per capita or life expectancy could transform peoples' outlook on life. Communities with higher income and better health are less concerned with struggles of daily survival. Self-expression values, which are the bedrocks of democracy, are more likely to flourish in these societies (Inglehart and Welzel, 2005). Also, more investment in human and physical capital could take place because of a longer lifespan. Working-class people could better organize and solve their collective action problem and modify the power structure of society toward a more responsive and representative system (Jacobsen, 2015).

*4.2. IVBMA results - developing countries*

Authoritarianism is the prevalent form of government in many developing countries. To have a more accurate understanding of the underlying causes of democracy, we focus on developing countries to see what determines democracy in this group of countries. We loosely consider non-OECD nations as developing. Using the data of 80 non-OECD countries[15], we repeat our IVBMA analysis.

Our findings highlight arable land as the most important determinant of democracy with a PIP of 0.919. Its positive posterior mean reminds that more arable land strengthens democracy in developing countries. We invite readers to pay attention to the arable land's posterior distribution in Figure 2. The compact distribution of arable land's posterior coefficients (in both the full sample and the non-OECD sample) indicates its low parameter variance. It is interesting to compare arable land's posterior distribution with that of the GDP per capita and see how densely the posterior coefficients are congregated around the posterior mean.

None of the youth population, life expectancy, and GDP per capita variables from our main IVBMA model turn out to be important for the non-OECD countries, in terms of affecting democracy. However, our IVBMA results show that state fragility with a PIP of 0.779 is a strong predictor of democracy in developing countries. The negative sign of its posterior mean tells that increasing state fragility decreases democracy.

Lipset (1959, 1995) give a word of warning about the pitfalls of institutionalizing democracy in developing countries. According to him, political stability is necessary for the consolidation of democracy. Political stability relies on a government's legitimacy and effectiveness. Legitimacy is "an accepted systemic title to rule," a widespread belief among the populace and the elite that the current political system is the best one to administer the society. A democratic government gains legitimacy through efficient performance in economic and political arenas. A government's success in providing relative plenty for its citizens shows its economic effectiveness and bolsters its legitimacy and stability. Safeguarding the interests of various powerful groups, such as military, adversary political groups, and business leaders constitute another element of legitimacy and stability. Healing social divisions and historical cleavages and obtaining the allegiance of the conservative or clerical groups are important for maintaining political effectiveness, legitimacy and the stability of a government. A stable democracy defines entry paths to the political process for social groups and new strata in order to discourage them from using forceful and violent means. Berg-Schlosser (2008) opines that broad acceptance by the people is key to the survival of democracies in Africa. Our IVBMA outcomes unfold that developing countries have not performed well regarding these issues and that democracy is threatened by fragile legitimacy and low effectiveness in these countries.

---

15 . We consider Latvia and Lithuania as non-OECD because they were not OECD members during 1991-2010.

Table 3 displays the data of the arable land, youth population, life expectancy, GDP per capita, and state fragility for the fifteen most and fifteen least democratic countries in our sample.

[Figure 2]

V. CONCLUSIONS

What are the true determinants of democracy? Literature gives no absolute answer to this question. In this article, we have tried to take stock of democracy literature by investigating the robustness of 42 covariates of democracy. We employ the Instrumental Variable Bayesian Model Averaging (IVBMA) approach and use the data of 111 countries during 2001-2010. We tackle endogeneity, a serious problem neglected by similar studies, by instrumenting the potential endogenous explanatory variables with their respective lagged values from 1991 to 2000.

For the full 111-country sample, our IVBMA outcomes shed light on the role of the arable land, youth population, life expectancy, and GDP per capita as the most influential predictors of democracy. Arable land with a PIP of 0.961 is at the top of the list, then stands youth population (PIP: 0.893), life expectancy (PIP: 0.839), and GDP per capita (PIP: 758). All have a direct relationship with democracy (i.e. their increase strengthen democracy), except the youth population, the increase of which erodes democracy. We believe the identification of arable land as the most robust determinant of democracy supports Crenshaw's (1995) proto-modernization hypothesis. He maintains that a country's ability to adopt democratic forms of government is affected by its ecological and historical heritages. Countries that experienced sophisticated rich agrarian societies in their histories are more likely to harbor democracy. Acemoglu et al. (2001, 2002, 2009) and Engerman and Sokoloff (2002) lend valuable insight into the lasting effects of geographical features and historical events in determining the countries' institutional development paths. Regarding the youth population, our findings parallel studies that warn about the adverse effects of large youth populations on democracy (Cincotta, 2008a, 2008b; Weber, 2013). The emergence of life expectancy and GDP per capita is evidence in favor of modernization theory and indicates the positive impact of economic development in spreading democracy (Wucherpfennig and Deutsch, 2009; Jacobsen, 2015).

We conduct a separate IVBMA analysis for 80 non-OECD countries. Arable land holds its position as the most persistent determinant of democracy even in the context of developing countries (PIP: 0.919) and retains its positive relationship with democracy. The other three variables (i.e. youth population, life expectancy, GDP per capita) lose their significance. For our subsample of developing countries, the IVBMA model selects state fragility (PIP: 0.779) as an important independent variable in explaining democracy levels. And in line with Lipset's (1959, 1995) arguments, state fragility turns out to be inimical to democracy.

Our study selects two different types of independent variables as the underlying causes of democracy. One group is relatively unchangeable or very hard to alter. Arable land belongs to this group. However, this does not mean that nothing could be done to promote democracy. According to our results, governments and international organizations could promote democracy by putting effective policies in place to modify societies' age structures and control the youth population size. Governments should embark on programs to reduce their fragility, by enhancing their efficiency and acquiring legitimacy. Economic development and increasing life expectancy and GDP per capita are other requisites of democracy.

Our study has its caveats and we want to point out two of these shortcomings which we believe are the most important ones. First, while we tried to be as inclusive as possible, our sample does not include some important authoritarian states such as Saudi Arabia and the Democratic Republic of the Congo. Second, although our approach to select the instruments is based on literature, our instruments might not fully satisfy the exclusion restriction.

We believe it is necessary to repeat this kind of analysis and test the robustness of our findings. Possible avenues for future research are: (a) replicate this study with other democracy measures (e.g. Polity IV or Freedom House indices), (b) using other types of instruments with more confidence in their ability to satisfy the exclusion restriction, (c) moving away from cross-section analysis and conduct panel data research since the latter enables investigating democratic transitions, (d) incorporating spatial dependencies into the instrumental variable models, (e) using non-linear econometric methodologies to account for the non-linear dynamics between democracy and its determinants, and (f) a combination of a, b, c, d, and e.

DISCLOSURE STATEMENT

We wish to confirm that there are no known conflicts of interest associated with this publication and there has been no significant financial support for this work that could have influenced its outcome.

[Table 3]

[Table 4]

[Table 5]

[Table 6]

Table 1. Empirical studies on the determinants of democracy.

| Paper | Dependent Variables | Method | Main Conclusions |
|---|---|---|---|
| Masi and Ricciuti (2019) | Polity2, VC, VP, VID | Synthetic Control Method | Oil discovery (-) |
| Oberdabernig et al. (2018) | Polity2, CL, PR, VID | Spatial Bayesian Model Averaging | Spatial spillovers are significant. Muslim religion (-), population (-), trade volumes (-), English language (+), natural resource rents (-), GDP per capita (+), dummy for MENA countries (-), armed conflicts (-) |
| Gerring et al. (2018) | Polity2, VC, CL, PR, IC, UDS | OLS, Instrumental variables | Ethno-linguistic diversity (+), religious diversity (-) |
| Wilson and Dyson (2017) | VID, Democ | OLS | Fertility rates (-) |
| Balaev (2014) | Polity2, Polyarchy, Freedom in the World | OLS | Economic development (+), education (+), gender equality (+) |
| Benhabib et al. (2013) | Polity2, VID, PR | OLS, Two-sided Tobit | Income (+) |
| Gassebner et al. (2013) | Przeworski et al. (2000) dichotomous variable | Extreme Bounds Analysis | Transition to democracy: GDP growth (-), past transitions: (+), OECD membership (+), fuel exporter (-), Muslim country (-), Survival of democracy: GDP per capita (+), past transitions (-), former military leader as the chief executive (-), democratic neighbors (+) |
| Hegre et al. (2012) | A dichotomous variable based on Polity IV | Dynamic logit models | Transition to democracy: natural resources (-), economic growth (-), multi-party (+), Survival of democracy: GDP per capita (+), economic growth (+), natural resources (+) |
| Berg-Schlosser (2008) | A categorical variable based on VC, VP, CL, and PR | Step-wise regressions, Qualitative Comparative Analysis | Ethnic and religious cleavages (-), mineral resources (-), rule of law (+), control of corruption (+), regulatory quality (+) |

(+): positive effect on democracy. (-): negative effect on democracy. Polity2: polity2 index from Polity IV dataset. VC: Vanhanen's competition index. VP: Vanhanen's participation index. VID: Vanhanen's index of democratization. CL: civil liberties index from Freedom House. PR: political rights index from Freedom House. IC: index of contestation by Coppedge et al. (2008). UDS: Unified Democracy Score from Pemstein et al. (2010). Democ: *Democ* measure from Polity IV. Polyarchy: by Vanhanen (2000). Freedom in the World: by Freedom House.

Table 2. Estimation results for dependent variable *democracy*.

| Variables | BMA results Full sample (111 countries) | | | Main IVBMA results Full sample (111 countries) | | | IVBMA results 80 non-OECD countries | | |
|---|---|---|---|---|---|---|---|---|---|
| | PIP | Post Mean | Post SD | PIP | Post Mean | Post SD | PIP | Post Mean | Post SD |
| Arable land | 0.423 | 0.054 | 0.071 | **0.961** | **0.156** | **0.054** | **0.919** | **0.178** | **0.076** |
| Youth population * | 0.244 | -0.168 | 0.336 | **0.893** | **-0.687** | **0.353** | 0.301 | -0.022 | 0.225 |
| Life expectancy * | 0.281 | 0.096 | 0.171 | **0.839** | **0.376** | **0.216** | 0.374 | 0.087 | 0.144 |
| GDP pc * | 0.339 | 2.596 | 4.046 | **0.758** | **1.118** | **1.042** | 0.609 | 0.574 | 0.909 |
| Muslim population | **0.952** | **-11.614** | **3.927** | 0.731 | -1.004 | 1.011 | 0.689 | -0.848 | 0.970 |
| State fragility * | 0.065 | -0.020 | 0.114 | 0.471 | -0.207 | 0.306 | **0.779** | **-0.510** | **0.368** |
| Primary education * | 0.038 | -0.037 | 0.276 | 0.603 | 0.503 | 0.660 | 0.451 | 0.156 | 0.515 |
| Colony: FR | 0.036 | -0.075 | 0.579 | 0.596 | -0.535 | 0.813 | 0.558 | -0.402 | 0.765 |
| Socialist legal origin | **0.841** | **-7.433** | **3.972** | 0.594 | -0.524 | 0.837 | 0.497 | -0.135 | 0.677 |
| MENA | 0.063 | -0.307 | 1.458 | 0.592 | -0.526 | 0.857 | 0.636 | -0.673 | 0.899 |
| Military leader | 0.036 | -0.065 | 0.457 | 0.585 | -0.501 | 0.767 | 0.605 | -0.562 | 0.825 |
| Latin America & Caribbean | 0.174 | 0.866 | 2.129 | 0.553 | 0.402 | 0.756 | 0.598 | 0.532 | 0.840 |
| East Asia & Pacific | 0.119 | -0.514 | 1.610 | 0.544 | -0.368 | 0.744 | 0.502 | -0.185 | 0.686 |
| Europe & Central Asia | 0.655 | 4.717 | 3.876 | 0.530 | 0.320 | 0.725 | 0.540 | 0.332 | 0.746 |
| Religious fractionalization | 0.025 | -0.005 | 0.521 | 0.521 | 0.254 | 0.741 | 0.527 | 0.277 | 0.742 |
| Colony: SP | 0.053 | 0.173 | 0.965 | 0.515 | 0.256 | 0.711 | 0.580 | 0.478 | 0.804 |
| Herfindahl index | 0.024 | -0.009 | 0.575 | 0.511 | -0.209 | 0.717 | 0.517 | -0.232 | 0.719 |
| Sub-Saharan Africa | 0.137 | -0.636 | 1.828 | 0.502 | 0.114 | 0.693 | 0.489 | -0.032 | 0.666 |
| Latitude | 0.075 | 0.713 | 3.018 | 0.497 | 0.076 | 0.702 | 0.495 | 0.026 | 0.705 |
| South Asia | 0.074 | 0.374 | 1.598 | 0.496 | 0.086 | 0.689 | 0.494 | 0.055 | 0.694 |
| Language fractionalization | 0.029 | 0.056 | 0.563 | 0.495 | 0.149 | 0.668 | 0.488 | 0.057 | 0.664 |
| North America | 0.025 | 0.042 | 0.813 | 0.495 | -0.026 | 0.696 | 0.497 | 0.000 | 0.708 |
| Language: SP | 0.039 | 0.086 | 0.656 | 0.492 | 0.176 | 0.648 | 0.527 | 0.280 | 0.740 |
| Ethnic fractionalization | 0.029 | 0.066 | 0.648 | 0.489 | 0.081 | 0.681 | 0.493 | 0.032 | 0.665 |
| Language: FR | 0.022 | -0.012 | 0.279 | 0.480 | -0.150 | 0.641 | 0.481 | -0.095 | 0.622 |
| Population * | 0.039 | -0.043 | 0.310 | 0.469 | -0.198 | 0.527 | 0.441 | 0.008 | 0.556 |
| British legal origin | 0.025 | 0.000 | 0.281 | 0.469 | 0.035 | 0.587 | 0.479 | 0.010 | 0.617 |
| Language: ENG | 0.024 | -0.004 | 0.265 | 0.466 | 0.062 | 0.588 | 0.474 | 0.010 | 0.615 |
| Colony: UK | 0.026 | -0.010 | 0.298 | 0.465 | -0.023 | 0.586 | 0.495 | -0.185 | 0.645 |
| Secondary education * | 0.090 | 0.115 | 0.425 | 0.458 | 0.206 | 0.480 | 0.595 | 0.494 | 0.691 |
| French legal origin | 0.028 | 0.007 | 0.338 | 0.457 | -0.007 | 0.559 | 0.496 | 0.183 | 0.681 |
| Fertility rate * | 0.050 | -0.057 | 0.386 | 0.430 | 0.082 | 0.485 | 0.434 | -0.091 | 0.484 |
| Social globalization * | 0.048 | -0.003 | 0.034 | 0.339 | 0.065 | 0.110 | 0.298 | 0.052 | 0.098 |
| FDI * | 0.022 | -0.001 | 0.023 | 0.283 | -0.097 | 0.256 | 0.284 | -0.067 | 0.213 |
| Urbanization rate * | 0.154 | 0.016 | 0.042 | 0.281 | 0.027 | 0.050 | 0.324 | 0.037 | 0.063 |
| Female labor force * | 0.033 | 0.001 | 0.016 | 0.238 | 0.021 | 0.043 | 0.114 | 0.008 | 0.027 |
| Infant mortality * | 0.174 | -0.021 | 0.051 | 0.148 | 0.014 | 0.041 | 0.080 | 0.003 | 0.024 |
| Economic globalization * | 0.033 | -0.001 | 0.012 | 0.139 | 0.018 | 0.062 | 0.129 | 0.012 | 0.047 |
| Gini index * | **0.841** | **-0.315** | **0.170** | 0.117 | 0.004 | 0.047 | 0.189 | 0.026 | 0.079 |
| **Natural resources** | 0.079 | -0.013 | 0.053 | 0.095 | -0.002 | 0.034 | 0.109 | -0.002 | 0.042 |
| **Agricultural employment *** | 0.276 | -0.036 | 0.064 | 0.085 | -0.003 | 0.026 | 0.075 | 0.000 | 0.024 |
| **Fuel exports** | 0.271 | -0.019 | 0.036 | 0.042 | -0.001 | 0.008 | 0.068 | -0.003 | 0.014 |

In each of the three models, bold numbers indicate variables with PIP > 0.75. A * represents a potentially endogenous variable. Endogenous variables are instrumented by their lagged values in the IVBMA models. PIP: posterior inclusion probability. We have reported the second stage PIPs for the IVBMA models. Post Mean: posterior mean. Post SD: posterior standard deviation.

Table 3. The 15 most and 15 least democratic countries among all 111 countries.

| Countries | Democracy | Arable land | Youth population | Life expectancy | GDP pc | State fragility |
|---|---|---|---|---|---|---|
| *15 most democratic countries* | | | | | | |
| Denmark | 44.34 | 55.01 | 24.63 | 77.86 | 4.77 | 0.00 |
| Netherlands | 41.53 | 30.33 | 25.23 | 79.47 | 4.69 | 0.00 |
| Switzerland | 40.04 | 10.29 | 25.45 | 81.29 | 4.85 | 1.00 |
| Sweden | 39.02 | 6.48 | 25.35 | 80.63 | 4.70 | 0.00 |
| Norway | 38.54 | 2.35 | 26.06 | 80.02 | 4.94 | 2.00 |
| Cyprus | 37.04 | 11.07 | 33.23 | 78.73 | 4.48 | 3.00 |
| Finland | 37.04 | 7.35 | 24.88 | 78.96 | 4.66 | 0.00 |
| Austria | 36.86 | 16.67 | 25.66 | 79.58 | 4.65 | 0.00 |
| Greece | 36.84 | 20.36 | 27.92 | 79.35 | 4.44 | 0.50 |
| Germany | 36.02 | 34.04 | 23.75 | 79.08 | 4.60 | 0.00 |
| Italy | 35.37 | 25.98 | 24.29 | 80.98 | 4.57 | 0.70 |
| Spain | 34.41 | 25.60 | 28.49 | 80.46 | 4.49 | 0.70 |
| United States | 33.71 | 18.02 | 27.52 | 77.64 | 4.68 | 1.70 |
| Israel | 33.60 | 14.57 | 31.02 | 80.38 | 4.45 | 8.50 |
| Australia | 33.24 | 6.06 | 28.14 | 80.81 | 4.69 | 1.70 |
| *Mean* | 37.17 | 18.95 | 26.77 | 79.68 | 4.64 | 1.32 |
| *15 least democratic countries* | | | | | | |
| Cambodia | 6.41 | 21.02 | 36.27 | 63.27 | 2.80 | 13.60 |
| Kazakhstan | 6.08 | 10.60 | 35.07 | 66.58 | 3.87 | 9.00 |
| Tanzania | 5.59 | 11.18 | 33.76 | 56.65 | 2.80 | 12.70 |
| Gambia | 5.51 | 33.21 | 33.99 | 58.05 | 2.73 | 13.60 |
| Vietnam | 5.42 | 20.74 | 37.26 | 74.33 | 3.02 | 8.20 |
| Cameroon | 5.16 | 12.66 | 34.50 | 53.09 | 3.10 | 17.30 |
| Tunisia | 4.56 | 17.84 | 36.54 | 74.20 | 3.56 | 7.80 |
| Morocco | 4.31 | 18.42 | 35.85 | 71.68 | 3.39 | 6.50 |
| Burkina Faso | 4.13 | 18.68 | 33.63 | 53.81 | 2.71 | 16.70 |
| Mali | 4.08 | 4.57 | 33.16 | 52.25 | 2.82 | 15.70 |
| Iran | 2.73 | 9.89 | 42.56 | 72.15 | 3.76 | 14.50 |
| Egypt | 2.28 | 2.79 | 36.37 | 69.56 | 3.35 | 13.00 |
| Jordan | 0.77 | 2.06 | 37.09 | 72.67 | 3.53 | 6.10 |
| China | 0.00 | 11.78 | 33.73 | 74.01 | 3.48 | 9.60 |
| Swaziland | 0.00 | 10.31 | 37.21 | 47.74 | 3.53 | 8.40 |
| *Mean* | 3.80 | 13.72 | 35.80 | 64.00 | 3.23 | 11.51 |
| *Sample Mean* | 20.49 | 16.33 | 31.29 | 71.84 | 3.94 | 6.42 |
| *Sample SD* | 17.17 | 11.42 | 5.21 | 10.38 | 0.78 | 6.00 |
| *Sample Median* | 19.83 | 13.62 | 33.20 | 75.99 | 4.15 | 6.30 |

Table 4. Summary statistics of full sample (111 countries).

| Variables | Mean | Median | SD | Min | Max | Correlation with democracy |
|---|---|---|---|---|---|---|
| Democracy | 19.12 | 19.55 | 11.16 | 0.00 | 44.34 | 1 |
| *Economic development* | | | | | | |
| GDP pc | 3.69 | 3.66 | 0.67 | 2.35 | 4.94 | 0.76 |
| Urbanization rate | 57.03 | 59.80 | 22.01 | 9.52 | 100.00 | 0.59 |
| Secondary education | 6.37 | 6.00 | 0.95 | 4.00 | 9.00 | 0.21 |
| Primary education | 5.62 | 6.00 | 0.93 | 3.00 | 8.00 | -0.13 |
| Infant mortality | 28.82 | 19.40 | 27.21 | 2.38 | 107.77 | -0.68 |
| Agricultural employment | 28.82 | 19.88 | 24.61 | 0.89 | 92.01 | -0.66 |
| FDI | 4.46 | 3.47 | 4.39 | -4.61 | 25.84 | 0.11 |
| Life expectancy | 69.41 | 72.40 | 9.59 | 45.10 | 82.19 | 0.68 |
| Gini index | 38.48 | 38.05 | 8.67 | 22.70 | 64.76 | -0.50 |
| Economic globalization | 56.62 | 55.63 | 15.81 | 22.75 | 97.92 | 0.41 |
| Social globalization | 53.98 | 54.80 | 21.98 | 11.78 | 95.78 | 0.73 |
| Natural resources | 6.10 | 2.44 | 7.80 | 0.00 | 32.97 | -0.42 |
| Fuel exports | 14.42 | 4.07 | 23.21 | 0.00 | 97.37 | -0.17 |
| *Institutional* | | | | | | |
| Population | 7.14 | 7.09 | 0.66 | 5.70 | 9.12 | -0.08 |
| Youth population | 32.47 | 33.36 | 3.96 | 23.75 | 42.56 | -0.72 |
| Fertility rate | 2.80 | 2.26 | 1.59 | 1.18 | 7.59 | -0.60 |
| Female labor force | 57.93 | 61.88 | 16.15 | 14.07 | 87.20 | 0.22 |
| State fragility | 7.82 | 8.20 | 6.05 | 0.00 | 20.10 | -0.74 |
| Military leader | 0.18 | 0.00 | 0.39 | 0.00 | 1.00 | -0.28 |
| *Cultural* | | | | | | |
| *Colonial heritage* | | | | | | |
| Colony: UK | 0.26 | 0.00 | 0.44 | 0.00 | 1.00 | -0.12 |
| Colony: FR | 0.14 | 0.00 | 0.34 | 0.00 | 1.00 | -0.41 |
| Colony: SP | 0.14 | 0.00 | 0.35 | 0.00 | 1.00 | 0.02 |
| Language: ENG | 0.26 | 0.00 | 0.44 | 0.00 | 1.00 | -0.09 |
| Language: FR | 0.15 | 0.00 | 0.36 | 0.00 | 1.00 | -0.16 |
| Language: SP | 0.15 | 0.00 | 0.36 | 0.00 | 1.00 | 0.08 |
| British legal origin | 0.27 | 0.00 | 0.45 | 0.00 | 1.00 | -0.08 |
| French legal origin | 0.42 | 0.00 | 0.50 | 0.00 | 1.00 | -0.20 |
| Socialist legal origin | 0.23 | 0.00 | 0.42 | 0.00 | 1.00 | 0.03 |
| *Fractionalization and religion* | | | | | | |
| Ethnic fractionalization | 0.43 | 0.42 | 0.26 | 0.00 | 0.93 | -0.48 |
| Language fractionalization | 0.38 | 0.33 | 0.29 | 0.00 | 0.92 | -0.40 |
| Religious fractionalization | 0.44 | 0.45 | 0.24 | 0.00 | 0.86 | 0.06 |
| Herfindahl index | 0.57 | 0.49 | 0.24 | 0.18 | 0.98 | 0.09 |
| Muslim population | 0.18 | 0.02 | 0.30 | 0.00 | 0.99 | -0.49 |
| *Geographical* | | | | | | |
| Arable land | 16.88 | 12.54 | 13.95 | 0.33 | 61.40 | 0.22 |
| **Latitude** | 0.32 | 0.33 | 0.20 | 0.01 | 0.71 | 0.57 |
| **North America** | 0.02 | 0.00 | 0.13 | 0.00 | 1.00 | 0.13 |
| **Latin America & Caribbean** | 0.17 | 0.00 | 0.38 | 0.00 | 1.00 | 0.04 |
| **Sub-Saharan Africa** | 0.23 | 0.00 | 0.43 | 0.00 | 1.00 | -0.50 |
| **East Asia & Pacific** | 0.12 | 0.00 | 0.32 | 0.00 | 1.00 | -0.06 |
| **MENA** | 0.07 | 0.00 | 0.26 | 0.00 | 1.00 | -0.23 |
| **Europe & Central Asia** | 0.34 | 0.00 | 0.48 | 0.00 | 1.00 | 0.56 |
| **South Asia** | 0.05 | 0.00 | 0.21 | 0.00 | 1.00 | -0.05 |
| *Instrumental variables* | | | | | | |
| **GDP pc (1991-2000)** | 3.57 | 3.58 | 0.67 | 2.27 | 4.86 | - |
| **Urbanization rate (1991-2000)** | 54.06 | 56.39 | 22.44 | 7.33 | 100.00 | - |
| **Secondary education (1991-2000)** | 6.40 | 6.00 | 0.90 | 4.00 | 9.00 | - |
| **Primary education (1991-2000)** | 5.50 | 6.00 | 1.00 | 3.00 | 7.00 | - |
| **Infant mortality (1991-2000)** | 40.52 | 29.14 | 35.96 | 3.97 | 138.05 | - |
| **Agricultural employment (1991-2000)** | 32.31 | 23.21 | 25.37 | 0.52 | 91.93 | - |
| **FDI (1991-2000)** | 2.66 | 1.99 | 2.74 | -4.86 | 15.88 | - |
| **Life expectancy (1991-2000)** | 66.73 | 69.06 | 9.76 | 43.74 | 79.95 | - |
| **Gini index (1991-2000)** | 39.61 | 39.35 | 9.95 | 21.74 | 65.91 | - |
| **Economic globalization (1991-2000)** | 47.35 | 45.75 | 14.82 | 19.14 | 93.76 | - |
| **Social globalization (1991-2000)** | 44.20 | 44.19 | 22.74 | 6.63 | 88.97 | - |
| **Population (1991-2000)** | 7.08 | 7.01 | 0.66 | 5.65 | 9.08 | - |
| **Youth population (1991-2000)** | 32.55 | 32.73 | 2.94 | 26.37 | 37.60 | - |
| **Fertility rate (1991-2000)** | 3.22 | 2.70 | 1.78 | 1.21 | 7.72 | - |
| **Female labor force (1991-2000)** | 56.08 | 59.03 | 16.96 | 11.73 | 88.73 | - |
| **State fragility (1995-2000)** | 9.05 | 9.17 | 6.42 | 0.00 | 23.33 | - |

All variables represent averages from annual observations between 2001 and 2010 unless stated otherwise. The *Military leader* variable is computed based on annual data from 2001-2008.

Instrumental variables are averages of annual data from 1991-2000 with an exception. The *State fragility* is instrumented using annual data from 1995-2000.

Table 5. Data descriptions and sources.

| Variables | Source | Description |
|---|---|---|
| **Democracy** | Vanhanen (2016) | Vanhanen's Index of Democratization (the higher the more democratic) |
| *Economic development* | | |
| **GDP pc** | World Bank | Log (GDP per capita (constant 2010 US$)) |
| **Urbanization rate** | World Bank | Urban population (% of total) |
| **Secondary education** | World Bank | Secondary education, duration (years) |
| **Primary education** | World Bank | Primary education, duration (years) |
| **Infant mortality** | World Bank | Mortality rate, infant (per 1,000 live births) |
| **Agricultural employment** | World Bank | Employment in agriculture (% of total employment) (modeled ILO estimate) |
| **FDI** | World Bank | Foreign direct investment, net inflows (% of GDP) |
| **Life expectancy** | World Bank | Life expectancy at birth, total (years) |
| **Gini index** | Solt (2019) | Gini index of inequality in equivalised (square root scale) household disposable income (the higher the more inequality) |
| **Economic globalization** | Gygli et al. (2019) | KOF Globalisation Index, economic globalization, de facto |
| **Social globalization** | Gygli et al. (2019) | KOF Globalisation Index, social globalization, de facto |
| **Natural resources** | World Bank | Total natural resources rents (% of GDP) |
| **Fuel exports** | World Bank | Fuel exports (% of merchandise exports) |
| *Institutional* | | |
| **Population** | World Bank | Log (total population) |
| **Youth population** | United Nations (2017) | people aged 15-34 (% of total) |
| **Fertility rate** | World Bank | Fertility rate, total (births per woman) |
| **Female labor force** | World Bank | Labor force participation rate, female (% of female population ages 15-64) (modeled ILO estimate) |
| **State fragility** | Center for Systemic Peace | State Fragility Index (the higher the more instable) |
| **Military leader** | Cheibub et al. (2010) | Dummy variable coded 1 if the effective head is or ever was a member of the military by profession, 0 if civilian. |
| *Cultural* | | |
| *Colonial heritage* | | |
| **Colony: UK, Colony: FR, Colony: SP** | Mayer and Zignago (2011) | 3 dummies, coded 1 if the country was primarily colonized by United Kingdom, France or Spain |
| **Language: ENG, Language: FR, Language: SP** | Mayer and Zignago (2011) | 3 dummies, coded 1 if more than 9% of the population speak the indicated language |
| **British legal origin, French legal origin, Socialist legal origin** | La Porta et al. (1999) | 3 dummy variables for English common law, French, and socialist legal systems |
| *Fractionalization and religion* | | |
| **Ethnic fractionalization** | Alesina et al. (2003) | Ethnic fractionalization |
| **Language fractionalization** | Alesina et al. (2003) | Language fractionalization |
| **Religious fractionalization** | Alesina et al. (2003) | Religious fractionalization |
| **Herfindahl index** | Barro and McCleary (2003) | Herfindahl index of religion shares, including non-religion 1970 |
| **Muslim population** | Barro and McCleary (2003) | % Muslim population 1970 |
| *Geographical* | | |
| **Arable land** | World Bank | Arable land (% of land area) |
| **Latitude** | La Porta et al. (1999) | Abs(latitude of capital)/90 |
| **North America** | | |
| **Latin America & Caribbean** | | |
| **Sub-Saharan Africa** | | |
| **East Asia & Pacific** | | 7 dummies based on World Bank regional classifications |
| **MENA** | | |
| **Europe & Central Asia** | | |
| **South Asia** | | |

Table 6. Full sample (111 countries) and average democracy scores from 2001–2010.

| Country | Democracy | Country | Democracy | Country | Democracy |
|---|---|---|---|---|---|
| ***East Asia & Pacific (N=13)*** | | Portugal | 26.4 | ***Sub-Saharan Africa (N=26)*** | |
| Australia | 33.24 | Russia | 19.81 | Burundi | 8.89 |
| China | 0 | Slovakia | 26.42 | Burkina Faso | 4.13 |
| Indonesia | 22.31 | Slovenia | 28.01 | Central African Republic | 6.78 |
| Japan | 29.24 | Sweden | 39.02 | Cameroon | 5.16 |
| Cambodia | 6.41 | Turkey | 19.97 | Ethiopia | 7.6 |
| South Korea | 25.57 | Ukraine | 31.11 | Gabon | 6.43 |
| Mongolia | 18.14 | ***Latin America & Caribbean (N=19)*** | | Ghana | 17.75 |
| Malaysia | 13.02 | Argentina | 29.46 | Guinea | 6.43 |
| New Zealand | 30.68 | Bolivia | 19.55 | Gambia | 5.51 |
| Philippines | 22.16 | Brazil | 27.5 | Guinea-Bissau | 11.84 |
| Singapore | 6.55 | Chile | 21.39 | Kenya | 10.77 |
| Thailand | 12.32 | Colombia | 11.09 | Lesotho | 8.73 |
| Vietnam | 5.42 | Costa Rica | 18.64 | Madagascar | 9.86 |
| ***Europe & Central Asia (N=38)*** | | Ecuador | 19.4 | Mali | 4.08 |
| Albania | 23 | Guatemala | 8.54 | Mozambique | 7.37 |
| Armenia | 21.24 | Honduras | 13.6 | Mauritius | 23.76 |
| Austria | 36.86 | Jamaica | 13.78 | Malawi | 16.91 |
| Azerbaijan | 13.02 | Mexico | 22.15 | Namibia | 9.12 |
| Bulgaria | 30.64 | Nicaragua | 21.5 | Niger | 6.71 |
| Belarus | 13.33 | Panama | 24.63 | Nigeria | 9.96 |
| Switzerland | 40.04 | Peru | 20.85 | Senegal | 9.61 |
| Cyprus | 37.04 | Paraguay | 14.25 | Swaziland | 0 |
| Czechia | 32.64 | Suriname | 22.76 | Tanzania | 5.59 |
| Germany | 36.02 | Trinidad & Tobago | 25.55 | Uganda | 10.61 |
| Denmark | 44.34 | Uruguay | 31.56 | South Africa | 11.63 |
| Spain | 34.41 | Venezuela | 14.53 | Zambia | 10.94 |
| Estonia | 26.87 | ***MENA (N=8)*** | | | |
| Finland | 37.04 | Algeria | 10.6 | | |
| France | 29.39 | Egypt | 2.28 | | |
| United Kingdom | 28.48 | Iran | 2.73 | | |
| Georgia | 11.03 | Israel | 33.6 | | |
| Greece | 36.84 | Jordan | 0.77 | | |
| Croatia | 26.93 | Lebanon | 21.88 | | |
| Hungary | 28.51 | Morocco | 4.31 | | |
| Ireland | 32.67 | Tunisia | 4.56 | | |
| Italy | 35.37 | ***North America (N=2)*** | | | |
| Kazakhstan | 6.08 | Canada | 26.41 | | |
| Kyrgyzstan | 8.86 | United States | 33.71 | | |
| Lithuania | 25.82 | ***South Asia (N=5)*** | | | |
| Latvia | 28.81 | Bangladesh | 16.02 | | |

| | | | |
|---|---|---|---|
| Moldova | 16.49 | India | 22.73 |
| Macedonia | 21.96 | Sri Lanka | 23.63 |
| Netherlands | 41.53 | Nepal | 14.29 |
| Norway | 38.54 | Pakistan | 6.82 |
| Poland | 21.51 | | |

APPENDIX

*BMA results*

Columns 2-4 in Table 2 disclose the BMA results. Unlike the IVBMA models, the BMA analysis does not tackle endogeneity. We focus our attention on variables with PIPs greater than 0.75. There are three covariates of democracy that meet this criterion: Muslim population, socialist legal origin, and Gini index. Among 42 potential predictors of democracy, the Muslim population proves itself as the most influential variable with a PIP of 0.952. Socialist legal origin and Gini index are the next most relevant determinants of democracy with PIPs of 0.841[16]. These three variables have negative posterior means indicating an inverse association with democracy (i.e. democracy deteriorates as they increase).

Our BMA findings reinforce the prevailing viewpoints in the literature. With respect to the Islam-democracy nexus, our research outcomes corroborate previous studies which detected a negative relationship between these two phenomena (Fish, 2002; Gassebner et al., 2013; Oberdabernig et al., 2018). Mobarak (2005) and Dutt and Mobarak (2016) explain that Islam not only regulates spiritual lives of its adherents, it also governs the political activities of its followers with a comprehensive set of divine laws and guidelines. Thus, nothing much is left to be decided by humans. Concerning socialist legal origin, our paper verifies conclusions from La Porta et al. (1999), Eichengreen and Leblang (2008), Friedman (2009), and Andersen (2012). Regarding income inequality, our results are similar to those of Andersen (2012) and Jung and Sunde (2014). Widening inequality correlates with crumbling democracy.

We refrain from giving BMA results too much of a credit, for the most part, because they suffer from endogeneity.

---

16 . Their PIPs differ slightly. PIPs are similar here due to rounding the decimals.

Figure 1. Average democracy scores 2001-2010.

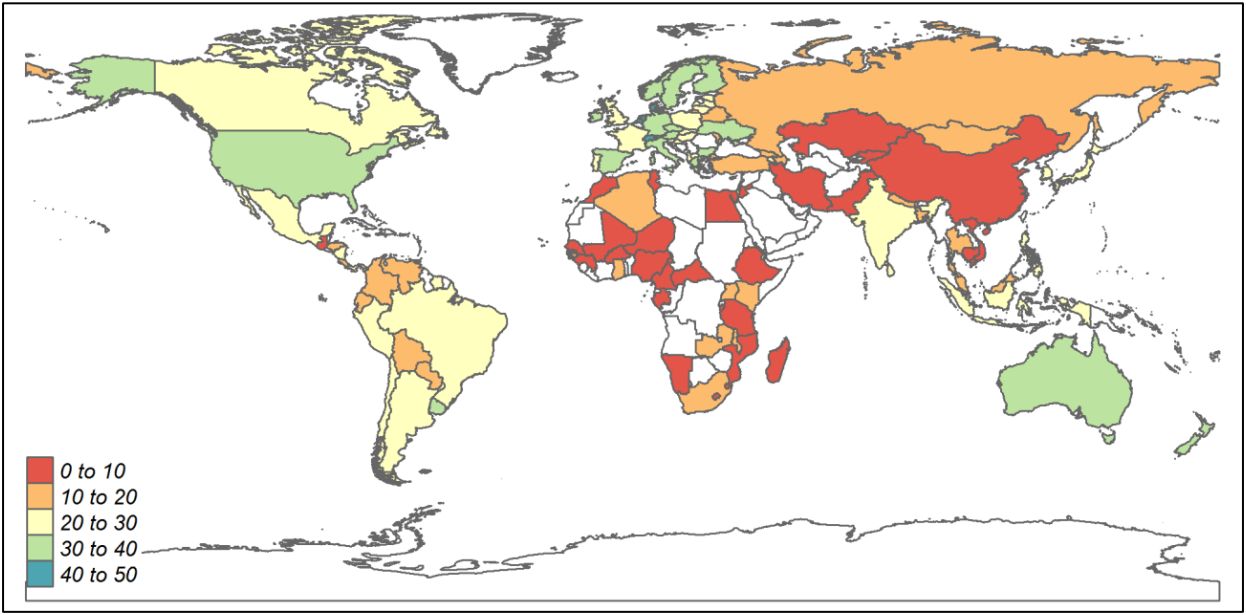

Higher scores indicate stronger democracy.

Figure 2. Posterior distributions of variables with PIP > 0.75 in the IVBMA models.

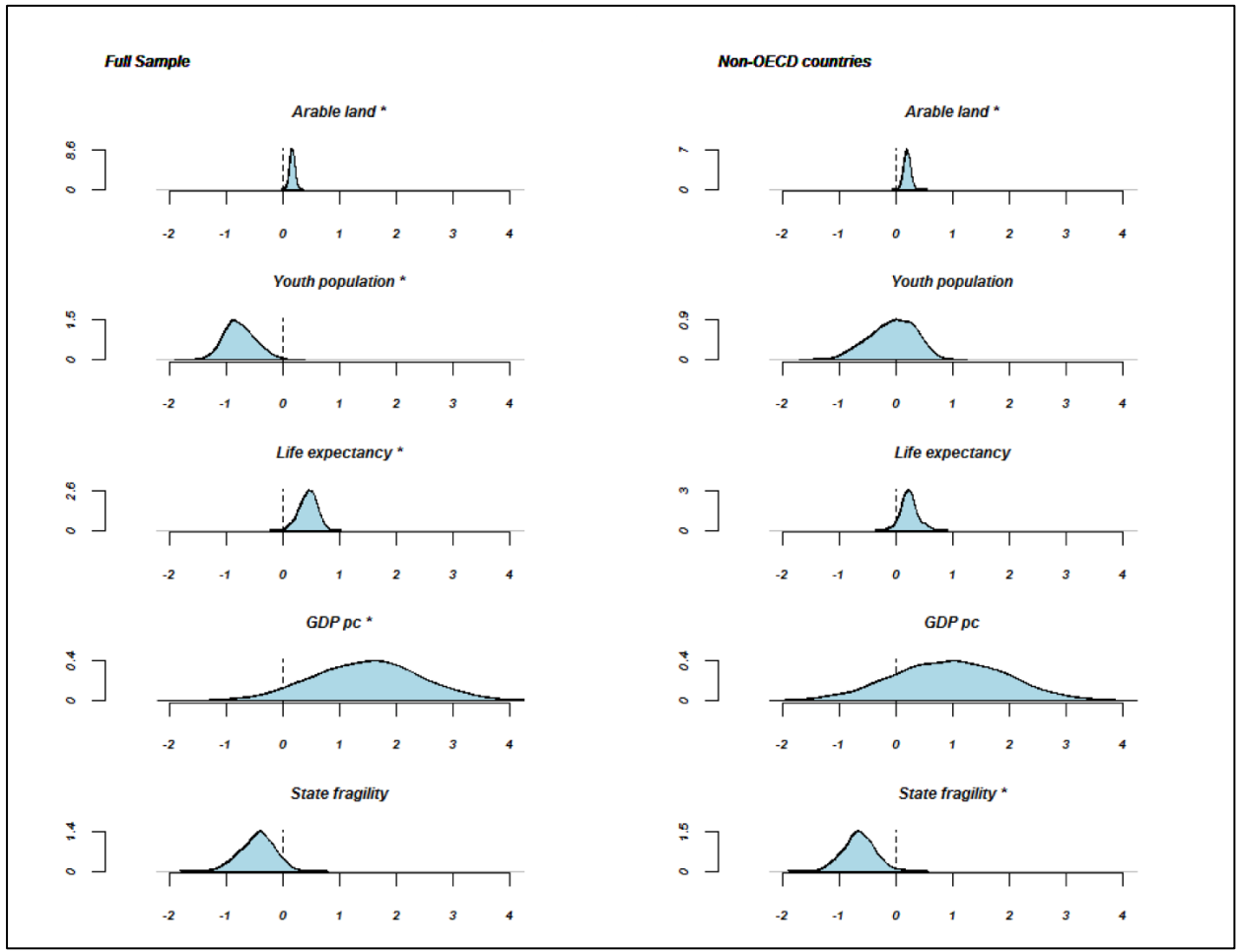

Variables with PIP > 0.75 are tagged with *. For each plot, X and Y axes present posterior coefficients and kernel density estimates respectively.